\documentclass[twocolumn,amssymb,amsmath,
nofootinbib,tightenlines,showpacs,floatfix,
superscriptaddress,aps,prb]{revtex4-1}
\usepackage{amsfonts}
\usepackage{txfonts} 
\usepackage{amsbsy}
\usepackage{bm}%
\usepackage[colorlinks=true,linkcolor=blue,filecolor=blue,menucolor=yellow,urlcolor=blue,citecolor=blue,anchorcolor=blue]{hyperref}
\usepackage{graphicx}
\usepackage{times} 
\usepackage{dcolumn}
\usepackage[style=base]{caption}
\usepackage[style=base,font=small]{caption} 
\usepackage[justification=justified]{subcaption}

\begin{document}

\title{Spin-split conductance and subgap peak in ferromagnet/superconductor 
spin valve heterostructures} 

\author{Evan Moen}
\email{moenx359@umn.edu}
\author{Oriol T. Valls}
\email{otvalls@umn.edu}
\altaffiliation{Also at Minnesota Supercomputer Institute, University of Minnesota,
Minneapolis, Minnesota 55455}
\affiliation{School of Physics and Astronomy, University of Minnesota, 
Minneapolis, Minnesota 55455}

\date{\today}


\begin{abstract}

We consider the separate spin channel contributions
to the charge conductance in 
superconducting/ferromagnetic spin valve $F_1/N/F_2/S$ structures. We find that
the up- and down-spin conductance contributions 
may have a very different behavior 
in the subgap bias region 
(i.e. there is 
a spin-split conductance). This leads to a subgap 
peak in the total
conductance. 
This peak behavior, which can be prominent also in $N/F/S$ systems,
is strongly dependent, in a periodic
way, on the thickness 
of the intermediate ferromagnetic  layer. 
We study this phenomenon using  a numerical
self consistent method, with additional insights gained from
an approximate analytic calculation for an infinite $N/F/S$ structure.
We study also the angular dependence on the relative magnetization angle between $F_1$ and
$F_2$ of both the spin-split and the total conductance. 
We do so for realistic material parameters and layer thicknesses
relevant to experimental studies on these devices. 
We also find that the spin-split conductance is
 highly dependent on the interfacial scattering
in these devices, and we carefully include these effects for realistic systems. 
A strong valve-effect is found for the angularly dependent
subgap peak conductance that is largely independent on the scattering and may prove useful in actual realizations of a superconducting spin valve device.

\end{abstract}


\maketitle

\section{Introduction} 

Interest in the use of superconductors in spintronic devices, 
such as spin valves\cite{tsyzu},  has been growing despite an
incomplete understanding of the underlying
basic physics of such devices. 
One of the main goals in 
incorporating
superconductors
in new spintronic devices is to reduce their energy 
consumption\cite{Bhatti2017}. Superconducting spintronics also garners intrinsic scientific interest 
because of the intricate proximity effects\cite{Buzdin2005} involved
between ferromagnets ($F$) and superconductors ($S$). Thus, superconducting spintronic devices, including spin valves $F_1/N/F_2/S$, have been 
proposed and studied\cite{igor,esch,zkhv,wvhg,kami,birge,transistor}. 
The $F_1/N/F_2$ layers compose the spin valve portion, and $N$ is a normal metal spacer, which is needed to enable control of the relative 
magnetization orientation of the two ferromagnets. 
One of the ways in which
superconducting spin valves differ from
ordinary  ones is that their transport properties are 
non-monotonic with the relative 
orientation of the ferromagnetic exchange fields\cite{wvhg,Moen2017,Moen2018}. 

The scientific interest in the unusual and useful
properties of $F/S$ structures arises  from
their antagonistic proximity effects. In ferromagnets, the exchange 
field works to split apart singlet Cooper pairs, 
favoring same-spin triplet states ($m_z=\pm1$). This leads to an
$F/S$ proximity effect that differs drastically
from that at $N/S$ interfaces. These proximity effects are very short ranged, 
and are oscillatory in position\cite{Buzdin1990,Halterman2002} due to the Cooper pairs acquiring a center of mass momentum\cite{demler}. 
This spatial dependence is important when analyzing the layer 
thickness dependence of thermodynamic properties of $F/S$ layered structures\cite{Buzdin2005}, 
including in ferromagnetic Josephson junctions\cite{ryaz2001,birge2018}, and transport properties in superconducting spin valve devices\cite{Bernardo2015,Moen2017}. 
Under certain conditions, it is possible for these structures to feature long range 
proximity effects. 
For a non-uniform magnetization texture such as in Holmium\cite{Bernardo2015, hoprl}, or for heterostructures with two or more non-collinear ferromagnetic exchange fields\cite{berg86,hbv,bvh,zdravkov2013} 
such as the $F_1/N/F_2/S$ case we study, the spin singlet of the s-wave superconducting Cooper pair correlations can effectively be rotated into the triplet states with $m_z=\pm1$. 
A non-collinear exchange field is necessary, 
because otherwise $S_z$ commutes with the Hamiltonian and only the $m_z=0$ 
triplet state can be induced. Due to the symmetry of the s-wave
Cooper pairs these triplet correlations in the ferromagnet are odd in time\cite{bvh} or, equivalently,
 in frequency\cite{berezinskii}. These $m_z=\pm1$ correlations are long ranged since they are not 
broken
apart by the exchange field\cite{bvermp,kalcheim,singh,ha2016,wvhg}. 
This yields a unique spin-valve effect in $F_1/N/F_2/S$ structures where the triplet correlations, induced by a non-collinear magnetization angle
between the ferromagnets, can lead to a non-monotonic angular dependence on the transport features\cite{wvhg,Moen2017}, 
as well as on the static physical properties such as
the transition temperature\cite{alejandro}. This angular dependence motivates much of our study into superconducting spin-valve structures. By considering the spin-dependent charge transport
in $F/S$ structures, we can gain further insight into this angular dependence of the  superconducting spin valve. We are also interested in how this affects
the spin dependence of the conductance in this structure, and compare it to the angularly independent $N/F/S$ system.

A charge current carries electrons and
holes in both the spin-up and spin-down states, which add up to
produce the total conductance of the circuit. 
When a device is spin polarized, 
we can see unusual changes to the conductance features arising from the
 difference in the spin channel transport, leading to each spin band having its own associated conductance that differs from that of the 
opposite spin channel\cite{johnson1998,johnson2001}. The separate spin
channel conductances can have 
features which diverge from those of the total conductance, 
which is why we collectively refer to the spin-polarized components of the conductance
as the spin-split conductance. 
In a superconducting/ferromagnetic heterostructure, 
the interplay between each spin channel in the ferromagnet with the energy gap of the superconductor can lead to dramatic effects in the overall conductance. 
At low bias, 
this interplay is mediated by Andreev reflections\cite{Andreev} in which an incoming electron is reflected as a hole and forms a Cooper pair in the superconductor. There are 
two types of Andreev reflection: ordinary Andreev reflection in which the electron/hole has opposite spin upon reflection, and anomalous Andreev reflection in which they have the same spin. 
It has been shown\cite{wvhg,linder2009,visani,niu,ji} that for $F/S$ interfaces, triplet proximity effects are correlated with anomalous Andreev reflection. 
Therefore, it is pertinent to consider these 
reflections when determining the spin-split conductance for  $N/F/S$ and 
or $F_1/N/F_2/S$ systems.

In previous work\cite{Moen2017}, we have noted that 
the conductance $G$ versus bias voltage $V$ curves
in $F_1/N/F_2/S$ structures
can exhibit a ``subgap'' peak structure 
below the critical bias. We explain in this paper that in general
the  low bias structure of $G$ in these devices is due to spin split
conductance behavior, and we study in some detail the features involved and what parameters influence them.
Specifically, we calculate the spin-split conductance of $N/F/S$ and $F_1/N/F_2/S$ heterostructures and verify
that the spin dependence of the conductance can lead to exotic behavior and unusual properties, e.g. in  
the layer thickness dependence in such structures\cite{Moen2017}. 
By studying the spin-split conductance, we can gain a deeper understanding of the full conductance features 
studied thus far. We  begin, in this work, 
 with a simple analytic model of an 
 $N/F/S$ structure with infinitely thick $N$ and $S$ layers and
 examine the thickness dependence of the ferromagnet for the spin-split 
conductance in an approximate non-self consistent approach. 
We then compare this model to a fully self-consistent numerical calculation for a finite nanoscale system. We then include a second 
ferromagnet to determine how the spin-split conductance can lead to the angular dependence in the total conductance. 
The numerical calculations are done by 
finding the self-consistent solution of the Bogoliubov de Gennes (BdG) equations\cite{degennes}, which determine the pair potential of the superconductor, with the 
proximity effects fully being taken into account. 
We then use a transfer matrix procedure within the Blonder-Tinkham-Klapwijk (BTK) method\cite{btk} to extract the conductance. 
We use realistic interfacial scattering strengths, 
as pertinent to good but imperfect experimental samples.  We find
that moderate interfacial
scattering actually enhance the spin-valve effects in some cases, as we discuss in our analysis and conclusions below.
We  also use layer thickness values relevant\cite{alejandro}
 to experimental studies of these devices. 
The exchange field of the ferromagnet and the coherence length of the superconductor are taken to ranges that correspond
 to the actual materials (such as Co and Nb)
used so that our work can be 
more easily compared with experimental results. We perform our calculations
in the low $T$  limit in order to best identify the spin-split conductance features that can be seen.

After having established  the properties of 
spin-split conductance via these analytical
and numerical methods, we then calculate the angular dependence 
of the conductance in the $F_1/N/F_2/S$ superconducting spin valve and establish how it is deeply related to the 
spin-split conductance. We also discuss how the interfacial scattering, which is an inevitable consequence of imperfect interfaces 
in fabricable devices, affect the spin-split features and the subgap peak conductance in these systems. From this study, we see a dramatic shift in the conductance for biases below 
the critical bas (CB) value, determined by the pair potential 
of the superconductor. This shift is oscillatory with the thickness of 
the $F_2$ layer, and it results in a conductance peak that occurs between the 
critical bias and zero bias. We find that  this subgap peak in conductance can have a large angular dependence, producing a significant valve-effect.
Thus, we hope  that our work will lead 
to a better understanding of these devices for future application and motivate additional theoretical and experimental work.

\begin{figure}
\includegraphics[width=0.45\textwidth] {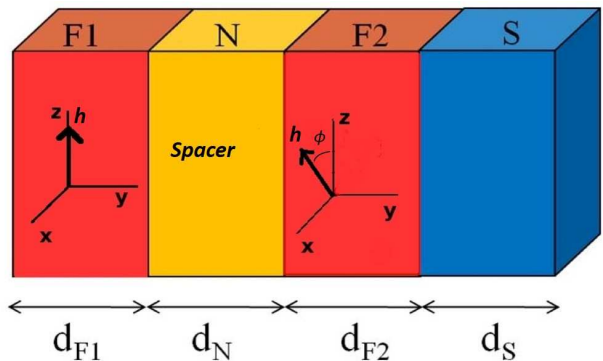} 
\caption{Sketch of the $F_1/N/F_2/S$ heterostructure with
the notation for thicknesses of the different layers indicated, The magnetizations
of the outer  magnetic layer $F_1$
is along the $z$ axis while that in
$F_2$ it is in the $x-z$ plane, forming
an angle $\phi$ with the $z$ axis, as indicated. The $y$ axis is 
normal to the layers. In the $N/F/S$ system, the first layer labeled $F_1$ is omitted.
This sketch is not to scale. 
}
\label{figure1}
\end{figure}

\section{Methods}
\label{methods}

\subsection{The basic equations}
\label{basic}

Since
the methods we use here are those of 
Refs.~[\onlinecite{wvhg,Moen2017,Moen2018}] we 
will merely, in this subsection, 
go over the main points to establish notation and to highlight certain
theoretical features that we wish to study.
Our focus 
is on the spin valve $F_1/N/F_2/S$ heterostructure,
and we will carry on the development in this context.  
The geometry of the valve structure 
is represented qualitatively in Fig.~\ref{figure1}.
The layers are assumed 
to be infinite in the transverse direction and the $y$-axis is 
taken normal to
the layers. 

The Hamiltonian of our system in terms of the usual creation and annihilation
operators is,
\begin{eqnarray}
\label{ham}
{\cal H}_{eff}&=&\int d^3r \left\{ \sum_{\alpha}
\psi_{\alpha}^{\dagger}\left(\mathbf{r}\right){\cal H}_0
\psi_{\alpha}\left(\mathbf{r}\right)\right.\nonumber \\
&+&\left.\frac{1}{2}\left[\sum_{\alpha,\:\beta}\left(i\sigma_y\right)_{\alpha\beta}
\Delta\left(\mathbf{r}\right)\psi_{\alpha}^{\dagger}
\left(\mathbf{r}\right)\psi_{\beta}^{\dagger}
\left(\mathbf{r}\right)+H.c.\right]\right.\nonumber \\
&-&\left.\sum_{\alpha,\:\beta}\psi_{\alpha}^{\dagger}
\left(\mathbf{r}\right)\left(\mathbf{h}\cdot\bm{\sigma}
\right)_{\alpha\beta}\psi_{\beta}\left(\mathbf{r}\right)\right\},
\end{eqnarray}
where $\bf{h}$ is the Stoner field. 
We choose the $z$ axis to be along the
internal field in the outer magnet $F_1$
(see Fig.~\ref{figure1}), 
while forming an angle $\phi$ with the $z$ axis
(and lying in the $x-z$ plane) inside the inner magnet $F_2$. 
It  vanishes in the superconductor $S$ and in 
the normal spacer $N$.  
${\cal H}_0$ is the single particle
Hamiltonian, in which we include the interfacial scattering.

Via a generalized Bogoliubov transformation, performed 
using the phase conventions
of Ref.~[\onlinecite{wvhg}] one can, in our quasi one dimensional
geometry, rewrite the eigenvalue
equation corresponding to the above Hamiltonian as:

\begin{align}
&\begin{pmatrix}
{\cal H}_0 -h_z&-h_x&0&\Delta(y) \\
-h_x&{\cal H}_0 +h_z&\Delta(y)&0 \\
0&\Delta(y)&-({\cal H}_0 -h_z)&-h_x \\
\Delta(y)&0&-h_x&-({\cal H}_0+h_z) \\
\end{pmatrix}
\begin{pmatrix}
u_{n\uparrow}\\u_{n\downarrow}\\v_{n\uparrow}\\v_{n\downarrow}
\end{pmatrix} \nonumber \\
&=\epsilon_n
\begin{pmatrix}
u_{n\uparrow}\\u_{n\downarrow}\\v_{n\uparrow}\\v_{n\downarrow}
\end{pmatrix}\label{bogo},
\end{align}
where $u_{n \sigma}$ and $v_{n\sigma}$ are the position (i.e. $y$) and 
spin dependent quasiparticle and quasihole amplitudes in the 
Bogoliubov transformation. We use natural units: $\hbar=k_B=1$.
The quasi one dimensional Hamiltonian
is ${{\cal H}_0}=-(1/2m)(d^2/dy^2)+\epsilon_\perp -E_F(y)+U(y)$ where
$\epsilon_\perp$ is the transverse energy. 
Thus, Eq.~(\ref{bogo}) is a set of decoupled equations, 
one for each $\epsilon_\perp$. 
$E_F(y)$ is the layer dependent band width: we take here $E_F(y)=E_{FS}$.
$U(y)$ is the interfacial scattering, which
we assume to be spin-independent and parametrize as
$U(y)=H_1\delta(y-d_{F1})+H_2\delta(y-d_{F1}-d_N)+H_3\delta(y- d_{F1}-d_N-d_{F2})$ 
with $H_i$ being numbered from the left-most 
to the right-most interfaces
(see  Fig.~\ref{figure1}). 
We define dimensionless barrier height 
parameters, $H_{Bi}\equiv H_i/v_F$, where $v_F$ is
the Fermi speed in $S$,  which characterize the
strength of the delta functions.

The calculation of the pair potential 
$\Delta(y)$ must\cite{wvhg,bagwell,sols2,sols,Moen2017}
in principle be performed self consistently, in order to
ensure that  
charge conservation\cite{baym} is preserved. The self
consistency condition is:

\begin{equation}
\label{del}
\Delta(y) = \frac{g(y)}{2}{\sum_n}^\prime
\bigl[u_{n\uparrow}(y)v_{n\downarrow}^{\ast}(y)+
u_{n\downarrow}(y)v_{n\uparrow}^{\ast}(y)\bigr]\tanh\left(\frac{\epsilon_n}{2T}\right), \,
\end{equation}
with the summation being over all eigenstates and 
the prime indicating that it  is limited to states
with eigenenergies within a cutoff $\omega_D$ from the Fermi level.
The quantity  $g$ is the superconducting coupling constant, assumed to be 
in the singlet channel, and to be
nonvanishing in $S$ only. To get a self-consistent result, we start with a suitable
initial choice of $\Delta(y)$ and iterate Eqs.~(\ref{bogo}) and (\ref{del})
until the input and output values of $\Delta(y)$ coincide. 
One can then derive the thermodynamic quantities
from the self-consistent  wavefunctions\cite{zkhv,bvh}, the results
of which have been found to be in 
agreement with experimental work\cite{alejandro}.

As mentioned in the Introduction 
we have considered also 
a simpler $N/F/S$ structure which can be solved  analytically
if one makes the additional approximation of treating the
pair potential non-self-consistently, with a constant value of $\Delta(y)$
in $S$.
This system
can be visualized by removing the left-most $F_1$ layer in Fig.~\ref{figure1}.
The analytic calculation is an approximation, 
done only as a means of comparison and of 
obtaining, as we shall see, some physical insights. A correct
 calculation requires a self-consistent approach.
Also, we consider only  one 
barrier at the $N/F$ interface
of dimensionless strength $H_B$.

\subsection{Transport}
\label{Transport}

We use the BTK formalism\cite{btk} to evaluate the conductance.
We first calculate the reflection and transmission amplitudes
for incoming electrons traveling perpendicular to the plane of our heterostructure
and then use the BTK method to extract the conductance, in Sec.~\ref{extraction},
which is given in terms of the spin dependent Andreev and
ordinary reflection amplitudes $a_{\sigma,\sigma^\prime}$
and $b_{\sigma,\sigma^\prime}$.
The incoming waves in terms of these amplitudes are compactly written
in the form
\begin{equation}
\label{f1waveup}
\Psi_{F1,\uparrow}\equiv\begin{pmatrix}e^{ik^+_{\uparrow1}y}+b_{\uparrow,\uparrow}e^{-ik^+_{\uparrow1}y}
\\b_{\downarrow,\uparrow}e^{-ik^+_{\downarrow1}y}
\\a_{\uparrow,\uparrow}e^{ik^-_{\uparrow1}y}
\\a_{\downarrow\uparrow}e^{ik^-_{\downarrow1}y}\end{pmatrix}\enspace\enspace
\end{equation}
for an incoming up spin particle in $F_1$, while for the down spin case one has:
\begin{equation}
\label{f1wavedown}
\Psi_{F1,\downarrow}\equiv\begin{pmatrix}b_{\uparrow,\downarrow}e^{-ik^+_{\uparrow1}y}
\\e^{ik^+_{\downarrow1}y}+b_{\downarrow,\downarrow}e^{-ik^+_{\downarrow1}y}
\\a_{\uparrow,\downarrow}e^{ik^-_{\uparrow1}y}
\\a_{\downarrow,\downarrow}e^{ik^-_{\downarrow1}y}\end{pmatrix}\enspace\enspace
\end{equation}
where the second spin index in the amplitudes $a_{\sigma,\sigma^\prime}$
and $b_{\sigma,\sigma^\prime}$ denotes the spin of the incoming particle, and
the first that of the reflected wave. The wavevectors are:
\begin{equation}
\label{wavevector}
k^{\pm}_{\sigma 1}=\left[(1-\eta_{\sigma}{h}_1)\pm{\epsilon}-{k_\perp^2}\right]^{1/2},
\end{equation}
with $\eta_\sigma \equiv 1(-1)$
for up (down) spins. $k_\perp$ is the length of the wavevector
corresponding to energy $\epsilon_\perp$. Here and below all 
wavevectors are in units of $k_{FS}$ and all
energies in terms of $E_{FS}$. 

The method to calculate these amplitudes has been discussed 
in previous work\cite{wvhg,Moen2017} for
the $F_1/N/F_2/S$ system and the self-consistent pair potential
and it would be superfluous to repeat the discussion here.

\subsection{Approximate analytic methods}
\label{NSCmethod}

If one foregoes treating the pair potential self-consistently,
it is possible to derive expressions 
for the relevant amplitudes which are in principle analytic,
although rather intricate. We do this here  
for  an  infinite $N/F/S$ heterostructure, where $N$ and $S$ are assumed
to be of infinite thickness, but $F$ is finite. 
The  expressions
for the incident waves, now impinging from $N$, are of the form given 
in Eqs.~(\ref{f1waveup}) and (\ref{f1wavedown}) but with a simplified
wavevector structure involving only the spin independent 
wavevectors  
\begin{equation}
\label{Nwavevector}
k^{\pm}_{N}=\left[1\pm{\epsilon}-{k_\perp^2}\right]^{1/2}.
\end{equation}
For the intermediate layers, the eigenfunctions contain both left- and right-moving plane
waves\cite{wvhg}. Thus the 
wavefunction for
intermediate $F$ layer has eight 
unknown coefficients. The $S$ layer
contains right-moving quasiparticles and left-moving quasiholes, with four\cite{wvhg} 
unknown coefficients. Our
plane wave expressions for $\Psi_F$ and $\Psi_S$ are exactly 
the same as Eqs.~(8) and (10) in Ref.~[\onlinecite{wvhg}] (with $\phi=0$ in the former) and we 
do not repeat them here. It should be recalled that in the equations for $\Psi_S$ 
the (non-self-consistent) pair potential is a constant, $\Delta_0$,
in $S$. 

We apply the continuity condition at each interface $\Psi_N(0)=\Psi_F(0)$, 
$\Psi_F(d_F)=\Psi_S(d_F)$, where for the infinite system, 
we conveniently choose the $N/F$ interface to be at $y=0$ and
the $F/S$ interface to be at $y=d_F$. The conditions on their derivatives are 
$\partial\Psi_N(0)/{\partial}y=\partial\Psi_F(0)/{\partial}y+2H_B\Psi_F(0)$ and 
similarly for the second interface. We can use a transfer matrix method to write these
as $8\times8$ matrices $\mathcal{M}_i$ multiplied by their respective 
vector of unknown 
coefficients $x_i$ for each layer $i$, as was explained in Ref.~[\onlinecite{wvhg}].
Then, $\mathcal{M}_{N}x_{N,\sigma}=\mathcal{M}_{F,l}x_F$ and
$\mathcal{M}_{F,r}x_F=\mathcal{M}_{S}x_S$, where $(l,r)$ denote that the wavefunctions are evaluated on the left or right side of the layer respectively and $\sigma$
denotes the spin of the incoming electron in the $N$ layer. By solving and eliminating
the intermediate layer coefficients, we find the eight total coefficients of both the $N$ and $S$ layer:
\begin{equation}
\label{vectorsol}
x_{N,\sigma}=\mathcal{M}^-1_N\mathcal{M}_{F,l}\mathcal{M}^-1_{F,r}\mathcal{M}_Sx_S.
\end{equation}
Solving these eight 
equations simultaneously for both spin-up and spin-down incoming electrons, we find the 
two sets of four reflection amplitudes $b_{\sigma,\sigma^\prime}$ and $a_{\sigma,\sigma^\prime}$, one set for each incoming spin state $\sigma^\prime$, 
which we use to calculate the conductance in Sec.~\ref{extraction}.

Thus, the calculation is formally
analytic. 
Although the full form solution for each 
 reflection amplitude can not be written in a compact manner,
 knowing the form of the plane
wave description lets us approximately determine the spatial dependence of the amplitudes. 
This spatial dependence comes from a combination of plane waves in $F$,
which are of the form $e^{ik^\pm_{\sigma}d_F}$, in which the wavevectors in the $F$ layer are defined by Eq.~(\ref{wavevector}).
In the zero bias limit, $\epsilon\rightarrow0$, 
we can express the wavevector
for the forward conductance ($k_\perp=0$) 
as $k_{\sigma}=\sqrt{1{\pm}h}$ in our units, where we have
dropped the particle/hole notation as these quantities are the same
at zero bias. Thus if we write one such combination of plane waves, e.g. $e^{i(k_\uparrow-k_\downarrow)}{\approx}e^{-ihd_F}$ 
to lowest order in $h$, we expect then to see a spatial periodicity with a wavelength 
$d$ such as $k_{FS}d=2\pi E_{FS}/h$ (in dimensionless
units $d=2\pi/h$) at zero bias. These are the same 
as the well known oscillations of the Cooper pair amplitudes within the ferromagnet\cite{demler}. Similarly,  
$e^{i(k_\uparrow+k_\downarrow)}{\approx}e^{-ik_Fd_F}$ means we can also 
 expect oscillations of wavelength $d=2\pi/k_{FS}$ or, 
in dimensionless units, simply $2\pi$.
In subsection \ref{extraction}, we will
use the absolute value squared of these amplitudes to calculate the conductance. Therefore, we expect all real coefficients with a $2\pi/h$ or $2\pi$ periodicity
in the amplitude to result in a conductance with periodicities proportional
to $\pi/h$ and $\pi$ respectively. 

\subsection{Extraction of the spin split conductance}
\label{extraction}

From the above results one can extract
the conductance using the BTK method\cite{btk}. 
The current is related to the applied bias
$V$ via the expression: 
\begin{equation}
\label{totalcurrent}
I(V)=\int G(\epsilon)\left[f\left(\epsilon-eV\right) 
-f\left(\epsilon\right)\right]d\epsilon,
\end{equation}
where $f$ is the Fermi function. 
The bias dependent tunneling conductance is
$G(V)=\partial I/{\partial V}$
which we evaluate
in the low-$T$ limit.
The conductance can be calculated using the reflection amplitudes
$a$ and $b$ described in the above subsections, for either the self-consistent or non-self-consistent results. 
Combining the conductance contribution from incoming spin-up and
spin-down electrons one has:

\begin{align}
\label{spinsplitcond}
&G(\epsilon)=\sum_\sigma P_\sigma G_{\sigma }(\epsilon) 
\\\nonumber
&=\sum_{\sigma}P_{\sigma}\left(1+\frac{k^-_{\uparrow 1}}{k^+_{\sigma 1}}|a_{\uparrow,\sigma }|^2
+\frac{k^-_{\downarrow 1}}{k^+_{\sigma 1}}|a_{\downarrow,\sigma }|^2
-\frac{k^+_{\uparrow 1}}{k^+_{\sigma 1}}|b_{\uparrow,\sigma }|^2
-\frac{k^+_{\downarrow 1}}{k^+_{\sigma 1}}|b_{\downarrow,\sigma }|^2\right), 
\end{align} 
where $G$ is given in natural units of conductance $(2{\pi}e^2/\hbar)$, and
 $\sigma$ denotes the spin of the incoming electron. 
 In Eq.~(\ref{spinsplitcond}),
$k^\pm_{\sigma 1}$ denotes the wavevector of the respective particle/hole in the first layer. In the
$N/F/S$ case described in Sec.~\ref{NSCmethod}, $k^\pm_{\sigma 1}=k^\pm_N$ for both spins, while in the
$F_1/N/F_2/S$ case $k^\pm_{\sigma 1}$ is given by Eq.~(\ref{wavevector}). The factors
$P_\sigma\equiv (1-h_1\eta_\sigma)/2$ are included to take into
account the different density of incoming spin up and spin down states 
in the $F_1$ layer for the
$F_1/N/F_2/S$ system. In the $N/F/S$ system, $P_\sigma=1/2$ denoting equal density.
The quantities $G_\sigma$ are the spin-up and spin-down conductances,  
which we collectively refer to as the spin-split conductance, since each component may drastically differ and ``split''
in behavior from that of the total conductance $G$.
The energy dependence of $G(\epsilon)$ arises from the applied bias voltage $V$.
It is customary and convenient
to measure this bias in terms of the dimensionless quantity 
$E\equiv eV/\Delta_0$ where $\Delta_0$ is the value of the order parameter in
bulk $S$  material. We will refer to the dimensionless bias dependent
conductance simply as $G(E)$.
We will refer to the spin-split conductance
$G_\sigma$ in a similar fashion.  In the $F_1/N/F_2/S$ spin valve structure, $G$ and $G_\sigma$ also depend 
on $\phi$. 

Generally, $G_\uparrow$ and $G_\downarrow$ will differ 
significantly, however they are related to each other by a rotation 
around the $y$ axis
in spin space. Using
the unitary transformation\cite{hbv} $U=e^{-\frac{i}{2}\theta\sigma_y}$ and taking the expectation value, we 
can define our spin-up and spin-down conductances, $G_{\sigma}(\theta)$, in a 
 basis rotated  from that of the $z$ axis,  
$G_{\sigma}(0)$, as:
\begin{subequations}
 \label{Gspineq}
\begin{align}
G_{\uparrow}(\theta)=\cos^2(\theta/2) G_{\uparrow}(0)+\sin^2(\theta/2) G_{\downarrow}(0)\\
G_{\downarrow}(\theta)=\sin^2(\theta/2) G_{\uparrow}(0)+\cos^2(\theta/2) G_{\downarrow}(0)
\end{align}
\end{subequations}

In the $N/F/S$ system the angle $\theta$ can be thought of as 
the angle $\phi$ between the field in $F$
and the $z$ axis, since this basis rotation 
is exactly the same as a rotation in $F$. However, this is not the case in the 
$F_1/N/F_2/S$ system when there is an 
actual angular mismatch. We can thus compare the change in the spin
split conductance due to the angular mismatch to that 
arising from pure rotation in basis. 

\begin{figure*}
\vspace*{-1.5cm}
\includegraphics[angle=-90,width=0.95\textwidth] {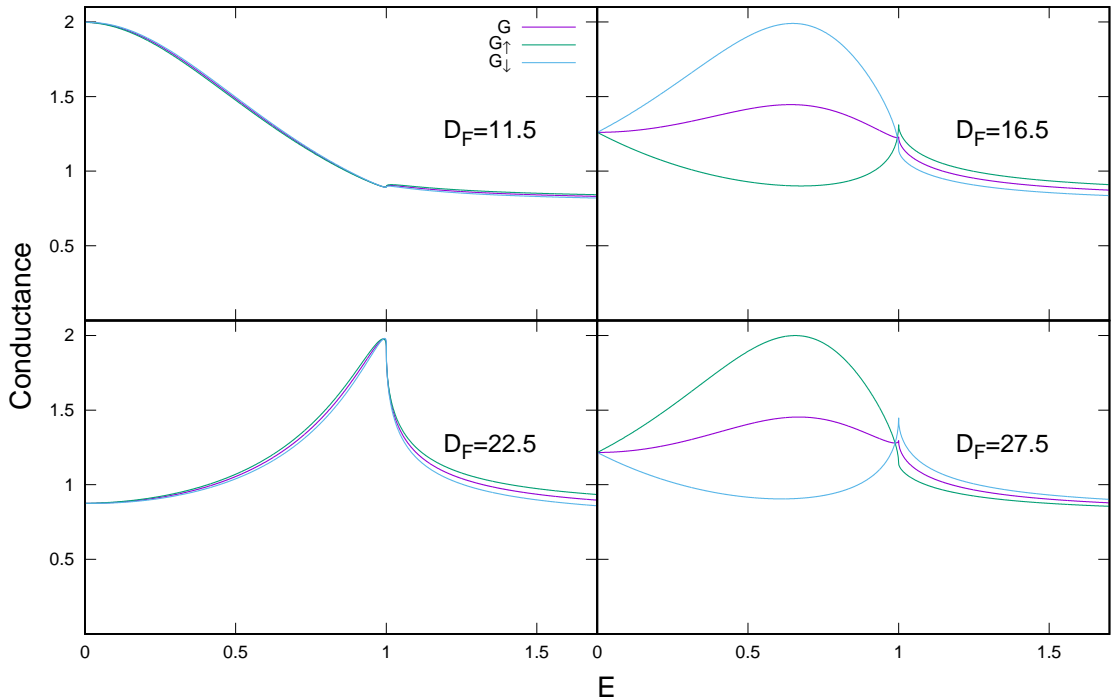} 
\vspace*{-0.9cm}
\caption{Spin-up, spin-down, and total conductances (see legend)
as a function of bias (E) 
in the infinite $N/F/S$ system. Conductances are calculated
using a non-self-consistent method. A single barrier with $H_B=0.5$ is  
at the $N/F$ interface. 
The four panels are for different values of the $F$ layer thickness, labeled $D_F$, which are
chosen at intervals of a quarter period ($\pi/2$ phase) of the spatial dependence.}
\label{NSCfig1}
\end{figure*}

\section{Results}
\label{results}

In this section we present our results for the spin-split conductance
defined by Eq.~(\ref{spinsplitcond}) as explained in the text below it.
We start (Sec.~\ref{analytic}),
by analyzing a simple,  
 $N/F/S$ system, with infinitely thick $S$ and $N$ layers, in a  non-self-consistent
manner, as derived in Sec.~\ref{NSCmethod}. In that case
the calculations can be performed analytically, and
the results, although quantitatively inaccurate,
illuminate a qualitative discussion that applies to all $F/S$ systems. 
We then move to the
 self-consistent approach (Sec.~\ref{FSspinsplit}) first briefly 
for a finite size $N/F/S$ 
system, so that we can gauge
the degree of reliability of the
analytic calculations, and then, in subsection~\ref{FFSspinsplit}, 
we consider the realistic superconducting spin valve $F_1/N/F_2/S$ 
system. For reasons
that will become clear below, we are 
particularly interested in how the conductance depends 
on the intermediate ferromagnetic layer thickness 
and on the interfacial scattering barriers, particularly that at the $N/F$ interface. 

In presenting our results we use dimensionless units: 
all lengths are in units 
of $k_{F S}$ and are denoted by capital letters such as $D_N$, $D_F$, 
and $D_S$. The bias voltage $E$ is in units of the bulk 
value of the pair potential, $\Delta_0$. As mentioned above, the conductance is in 
natural units $2 \pi e/\hbar$. 
Values of the dimensionless barrier parameters $H_B$ (introduced
above) greater than unity 
would begin to approach the tunneling limit, while zero represents 
a perfect interface. We also set any
 wavevector mismatch parameters to unity, subsuming their effects in 
the phenomenological $H_B$ values. 
This reduces the number 
of parameters governing the system. 
With a minor exception for illustrative purposes, 
we set the exchange field 
in all ferromagnets (which we assume
to be of the same material in the valve case) to be $h=0.145$ 
in our dimensionless units, where $h=1$ is the half-metallic limit, and
we set the 
coherence length $\Xi_0=115$ in our  length units. The values
 of $h$ and $\Xi_0$ chosen have been found to be suitable to 
the quantitative analysis of static quantities done on similar 
systems using Cobalt and Niobium\cite{alejandro}. 

We have found that the most
crucial geometrical parameter for our purposes is
the thickness of the intermediate $F$ layer
and consequently we examine, in each subsection,  the conductance 
dependence on this  layer thickness $D_F$, or $D_{F2}$, 
for the $N/F/S$ or the $F_1/N/F_2/S$ spin valve system respectively. 
We also examine the dependence on the barrier $H_B$ at 
the $N/F$ or $N/F_2$ interface, and also, in 
Sec.~\ref{FFSspinsplit},  on the   barrier strengths at 
all the interfaces. 
In Sec.~\ref{FFSspinsplit}, we also 
examine the dependence 
of the spin-split and total conductances on the mismatch angle $\phi$ 
of the exchange fields, $\mathbf{h_1}$ and $\mathbf{h_2}$.

\begin{figure*}
\vspace*{-1.5cm}
\includegraphics[angle=-90,width=0.95\textwidth] {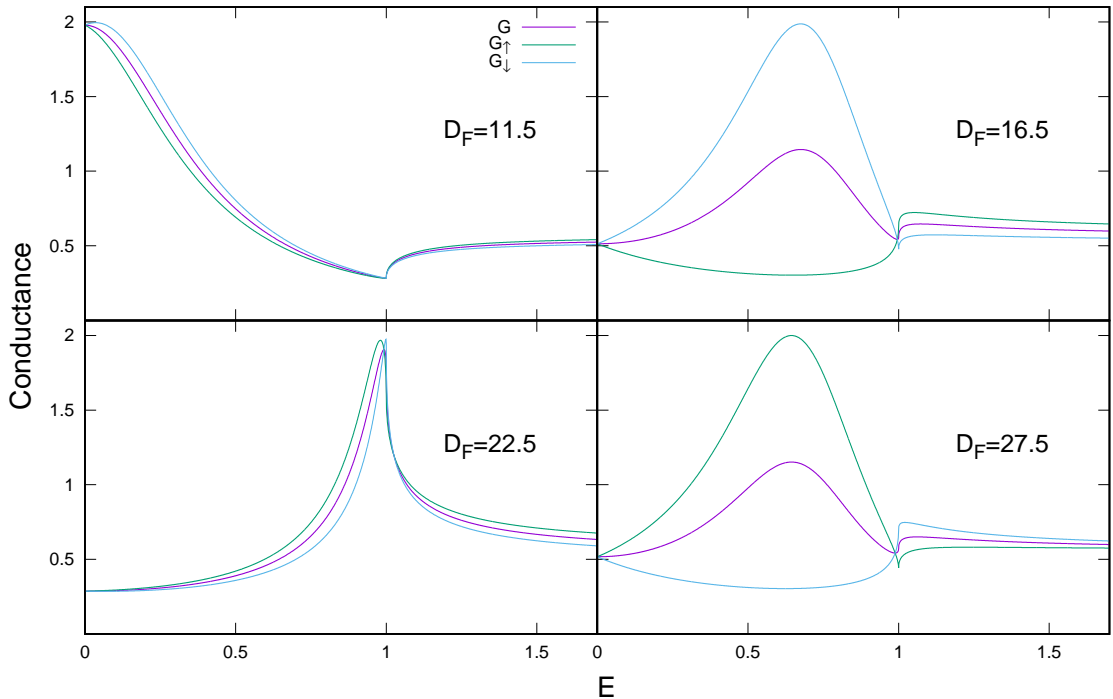} 
\vspace*{-0.9cm}
\caption{Spin-up, spin-down, and total conductance as a function of bias (E) in the infinite $N/F/S$ system. Conductance is calculated
as in Fig.~\ref{NSCfig1}. A barrier with $H_B=0.9$, significantly larger than that 
 in Fig.~\ref{NSCfig1} is  
at the $N/F$ interface. The four panels are arranged
as in the previous figure.}
\label{NSCfig2}
\end{figure*}

\subsection{N/F/S conductance: analytic results}
\label{analytic}

\begin{figure}
\vspace*{-.2cm}
\hspace*{-1.8cm}\includegraphics[angle=-90,width=.7\textwidth] {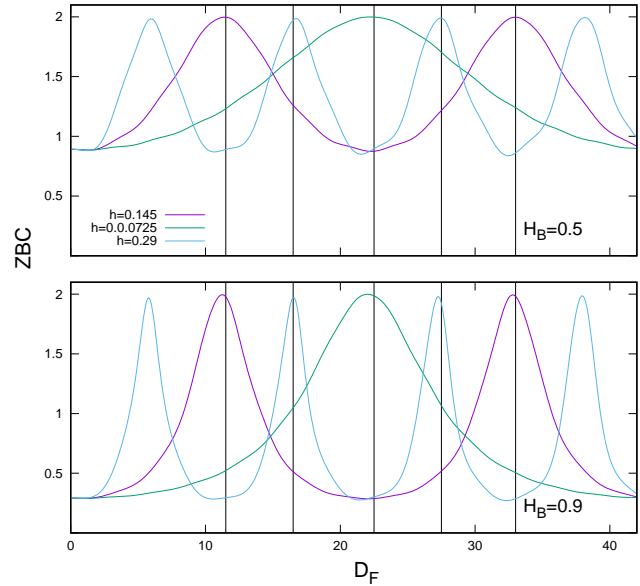} 
\vspace*{-0.9cm}
\caption{The zero bias conductance (ZBC) for an infinite $N/F/S$ system, 
as a function of $D_F$, for values of the
exchange field half of, equal to, and double $h=0.145$. The top and bottom panels 
have  barrier strengths of $H_B=0.5$ and
$H_B=0.9$ respectively, at the $N/F$ interface. We plot the $D_F$
 dependence for approximately two oscillation  periods at $h=0.145$.
We see a  wavelength of $\pi/h$ for all values of $h$.}
\label{NSCfig3}
\end{figure}

\begin{figure}
\vspace*{-.2cm}
\hspace*{-1.8cm}\includegraphics[angle=-90,width=0.7\textwidth] {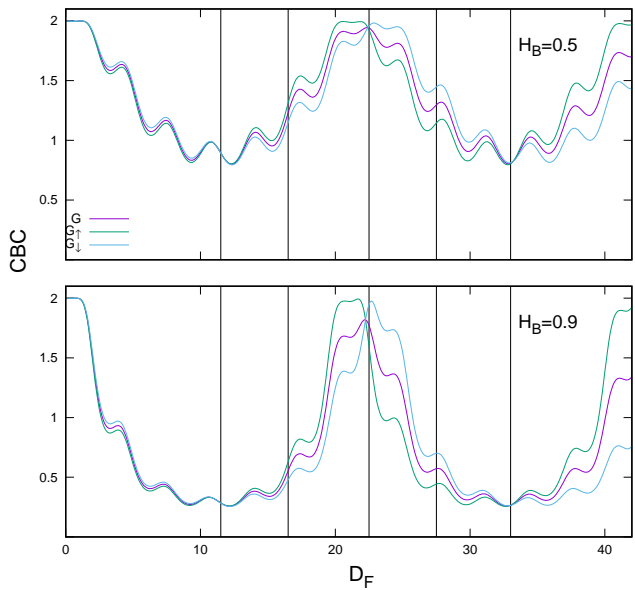} 
\vspace*{-0.9cm} 
\caption{The critical bias conductance (CBC) (total, sin-up and spin-down)
for an infinite $N/F/S$ system
as a function of $D_F$. 
The top panel and bottom panels have a barrier $H_B=0.5$ and
$H_B=0.9$ respectively, both at the $N/F$ interface. We plot the thickness dependence for approximately two periods of the $\pi/h$ oscillation.
We also see smaller, superimposed oscillations with periodicity of $\pi/k_F=\pi$.}
\label{NSCfig4}
\end{figure}

In this subsection we discuss the results of our analytic approach described in section \ref{NSCmethod}. 
To obtain analytic results, we have to abandon
self consistency, so the results are only approximate. We consider  
an infinite $N/F/S$ system, with finite, varying $D_F$ thickness
but infinite $D_N$ and $D_S$. This is worthwhile, however,
as from analytic results one can establish context and gain
a degree of physical insight that is difficult
to gather from 
our self-consistent numerical results discussed in the subsections 
below. 
The non-self-consistent results differ, of course, from
the correct self-consistent ones. One obvious
difference occurs near the critical bias (CB). 
For our analytic results the CB is always 
at $E=1$ since in the non-self-consistent case, $\Delta(Y)\equiv\Delta_0$ 
for all $Y$ in $S$. 
In Figs.~\ref{NSCfig1} and ~\ref{NSCfig2} 
we examine the spin-split conductance $G_\sigma$ (i.e. the
spin-up and spin-down components) and the total conductance, 
 $G$,  defined by Eq.~(\ref{spinsplitcond}),
as  functions of applied bias. 
In Figs.~\ref{NSCfig3} and \ref{NSCfig4} we plot the zero bias conductance (ZBC) and the critical 
bias conductance (CBC) respectively as  functions of the thickness of the ferromagnetic layer $D_F$. 

The results plotted in Fig.~\ref{NSCfig1} are for a moderately
strong barrier, $H_B=0.5$, which is a realistic value for
a good interface, located at the $N/F$ boundary. 
The reason for studying this barrier is that even a small amount 
of interfacial scattering 
allows for the formation of a prominent feature 
that we wish to study: the subgap conductance peak.  
This is 
a peak in the conductance occurring for specific thicknesses of $F$ 
at biases between zero and the 
critical bias (the ``subgap'' bias region). 
As we shall see, the spin-split conductance components can vary dramatically 
especially near the CB.
The total conductance is a combination of the components of the spin-split conductance.
In this single $F$ layer system, the total $G$ is 
simply the average of the up and down spin-band contributions. 
When the spin-up and spin-down conductances are split from one another, we see a peak in the total conductance where the two differ the most.
Examining the peak value of the conductance, 
we find a  periodic behavior 
with $D_F$, with a 
periodicity of $\pi/h\approx 22$ in our dimensionless units.
This can be traced, of course, to
the well-known periodicity\cite{demler} of the Cooper pair amplitudes,
reflected in the above given value, as has been discussed at the end of 
Sec.~\ref{NSCmethod}.  

The figure includes four panels, each
for a different value of $D_F$ within one cycle of 
this periodic behavior. In the first and third 
panels, which correspond to a $D_F$ difference
of about half a period, we see that the peak in the total conductance occurs at zero 
bias and at the critical bias respectively, while in
the similarly separated  second and fourth 
panels we see a subgap bias conductance peak.
This subgap peak in the conductivity is similar to those reported in Ref.~[\onlinecite{Bernardo2015}] for inhomogeneous $S/F$ structures,
where in their tunneling conductance measurements. They find
symmetrical, small peaks in the subgap region
of the density of states,  which they call the ``double-peak spectra'' and, for a subset of their samples, a single peak in the zero bias conductance, which they call 
the ``zero peak spectra''. Here, we will refer to these peaks as the subgap bias and the ZBC peaks respectively. We believe these observed tunneling conductance peaks
may be due to the spin-split conductance phenomenon we discuss below, with the ``zero peak spectra'' found for a small subset of their samples possibly being
due to small fluctuations in the sample layer thicknesses, at fractions of a nanometer.

The total
conductance peak moves away from zero bias in the first panel
to a finite subgap bias value in the second. Increasing $D_F$ further, 
the peak moves to the critical bias in  panel three, then 
returns in panel four
to the same subgap bias value as in panel two. It goes back 
to  zero bias, with a peak feature very
similar to that in panel one for $D_F=33$ (not shown):
at that point a whole period in $D_F$ has elapsed. 
In the first and third panels we see little 
difference between the subgap spin-up and spin-down conductances. 
On the other hand, we see a 
very large difference in the spin-split conductance for the second and fourth panels. 
In the second panel, 
the spin-down conductance has a large subgap peak, with $G$ reaching
a value of  $G=2$  before 
decreasing towards the CB, where 
there is a discontinuous change in slope (leading
to what we describe as a ``shoulder"). 
The spin-up conductance has the opposite behavior, with a dip in the subgap region 
that increases to a sharp cusp shaped peak at the CB. 
This spin-split conductance then yields a total 
$G$ with a local maximum at the 
spin-down conductance's maximum, which is also the spin-up 
conductance's minimum. 
In panel four, 
we see a very similar situation. However, the respective behaviors of the spin-up and spin-down conductances have reversed, 
with the spin-up conductance having an intermediate maximum and a 
shoulder critical bias feature, and the spin-down conductance having an 
intermediate minimum and cusp critical bias feature. In both 
of these panels, the CBC is also split 
between spin-up and spin-down, and the total conductance has a hybrid cusp-like behavior. There is then a crossing point, where each component (and the total conductance) meet, at a bias slightly below that of the CB. 

In all four panels  the ZBC is the
same for the spin-up and spin-down conductances, and consequentially 
for the total $G$. 
In ordinary Andreev reflection, a spin-up electron reflects 
into a spin-down hole, and vice versa. In the zero bias limit the 
electron and hole have equal energy. Thus, in the single $F$ layer case, the zero bias spin-up transmission amplitudes are the same 
as those for spin-down transmission, due to the symmetry of 
the electron/hole traveling in the spin-up/spin-down bands. 
We will see in Sec.~\ref{FFSspinsplit} that this is not the case when 
there is a second ferromagnetic layer.

In  Fig.~\ref{NSCfig2} we repeat the plots 
 in Fig.~\ref{NSCfig1} but 
for a stronger barrier, $H_B=0.9$. In all four panels we see in 
general a decrease in the conductance at all biases, with the
remarkable and interesting 
exception of the peak value of the
conductance, which remains high in  all cases. 
In panel one, for example, we see no decrease in the ZBC, and similarly in 
panel three for the CBC. This leads to a ``resonance'' feature 
similar to that discussed in Fig. 6 of
Ref.~[\onlinecite{Moen2017}], where the ZBC is independent on the barrier strength.
In panels two and four, we do see a moderate decrease in the 
average value of the total subgap conductance, but 
not in the maximum values
of the spin-split conductance. 
Instead, there is a decrease in the minimum of the 
opposite spin component, as well as a general decrease  
in the ZBC and CBC. This leads to a much  more pronounced
subgap peak conductance than in the $H_B=0.5$ case. This 
 feature is very resilient to high values of $H_B$, 
it  begins to deteriorate only  well
into the  tunneling limit. 
A low value for $H_B$ makes the peak
less obvious  
 as the subgap conductance increases
towards its maximum possible
value of $G=2$ and the difference between the spin-up and spin-down 
conductances decreases.
We have restricted our analysis
of this simplified model to the case of only one barrier at the $F/N$ interface. 
Below, in Sec.~\ref{FFSspinsplit}, we examine an $F/N/F/S$ system with barriers at each interface 
including $F/S$.

We now show specific
details of the $D_F$ periodicity. 
In Fig.~\ref{NSCfig3} we plot the zero bias conductance  
as a function of $D_F$ for $H_B=0.5$ (top panel) and $H_B=0.9$ 
(bottom panel), for $h=0.145$ (the 
only value we use in  
all of our  figures except this 
one) and for $h=0.0725$ and $h=0.29$,
half and double the original value. 
We do so to best demonstrate the dependence of
the periodicity on $h$. 
As mentioned above, 
the ZBC is equal for the spin-up, spin-down, and total conductances 
and therefore we only plot the total $G$. The 
four leftmost vertical lines 
in each plot are the values of $D_F$ used in Figs.~\ref{NSCfig1} and ~\ref{NSCfig2}, and the fifth is for  $D_F=33$ 
at which value  one full cycle is complete for
$h=0.145$. We can clearly see here the 
$\pi/h$ dependence of the wavelength of the oscillation. 
For a value double the original, the wavelength is halved, and vice versa. 
The oscillatory behavior looks 
very regular and fairly sinusoidal at $H_B=0.5$, 
except for some minor irregular variations which are more 
prominent for $h=0.29$. 
However, for the larger barrier value of the bottom panel, the oscillatory pattern is less sinusoidal, with a 
sharper dependence of the ZBC on $D_F$ at the ZBC maxima and a broadening of the ZBC minima. 
For the stronger barrier only a reduced range of thicknesses
have a ZBC peak conductance feature, which qualitatively agrees 
with what was found\cite{Bernardo2015} in non-homogenous $S/F$ structures. 
Near the vertical lines, we also see 
a slight change in the phase of the oscillation for
the stronger barrier. 
The periodic behavior breaks down for  values of $D_F$ of less
than a quarter-period, where the ZBC becomes constant and independent on $h$.

Following up on this we
plot, in Fig.~\ref{NSCfig4}, the critical bias conductance as a function of $D_F$, for both $H_B=0.5$ and $H_B=0.9$ (top and 
bottom panels respectively) at $h=0.145$. We do so for the spin-split
conductance components, which do not have the same CBC value,
as well as for the total conductance. We see the same 
overall periodic structure as 
in the ZBC, with a $\pi$ phase difference, since the CBC maxima occurs at the ZBC minima. 
There is also a minor oscillatory behavior with wavelength $\pi$ 
superimposed
on  the broader $\pi/h$ oscillations: this
is  unobservable in the ZBC. The $\pi$ oscillations are explained 
in Sec.~\ref{NSCmethod}. 
The spin-up and spin-down conductances cross over at 
the CBC maxima and they also converge at the CBC minima 
(where there are ZBC maxima). Between  nodal points there
is a difference 
 in the spin-split conductance components that reverses between a dominant 
spin-up or dominant spin-down conductance.  
The separation becomes greater as $D_F$ increases, or 
as the barrier strength increases. 
\begin{figure*}
\vspace*{-1.5cm}
\includegraphics[angle=-90,width=0.95\textwidth] {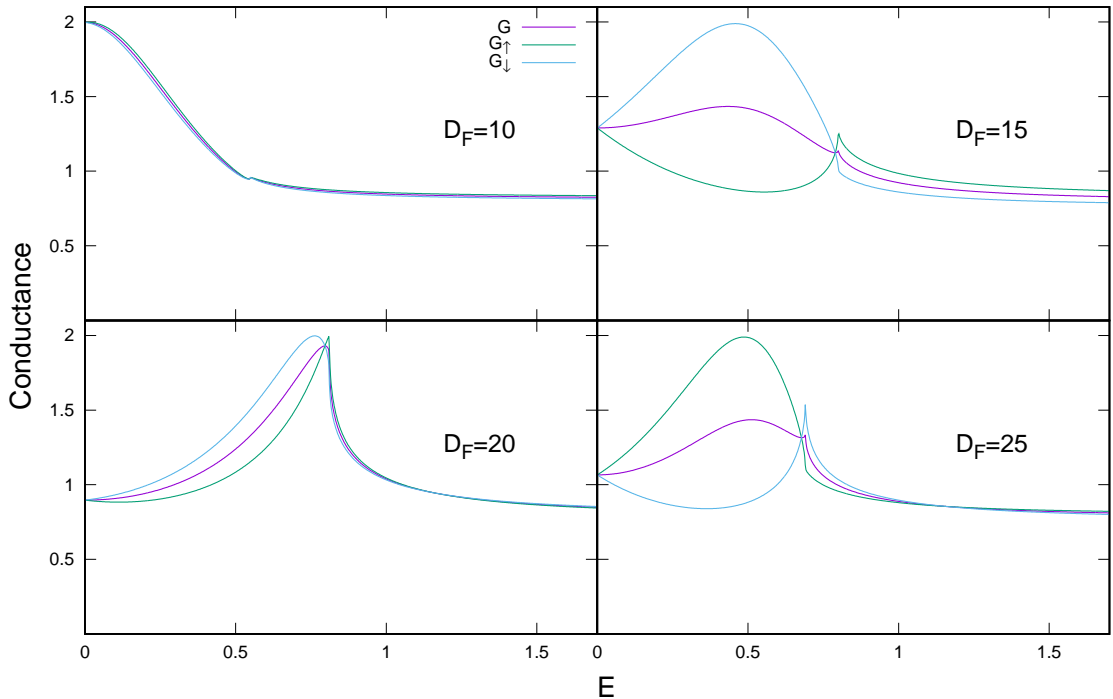} 
\vspace*{-0.9cm} 
\caption{Spin-up, spin-down, and total conductance as a function of bias (E) 
in the finite $N/F/S$ system with $D_N=90$ and $D_S=180$. Conductances are calculated
numerically using a self-consistent method. A barrier $H_B=0.5$ is  
at the $N/F$ interface only. The four panels are for different values of 
$D_F$, which are
plotted for intervals of a quarter period of the spatial dependence.}
\label{NFSfig1}
\end{figure*}

We have found, using approximate
analytic results,  a regular, periodic behavior in the conductance 
features as a function of the ferromagnetic layer thickness. 
We have also found a subgap bias conductance peak, 
the prominence of which
increases  with the strength
of the  scattering barrier at the $N/F$ interface. 
This peak is due to the splitting of 
the spin-up and spin-down conductances. 
This analysis will be helpful in interpreting the numerical results below.

\subsection{N/F/S spin-split conductance}
\label{FSspinsplit}


To make  the discussion of
our numerical spin valve results more understandable, 
we start with a brief discussion of a simpler finite  size, 
$N/F/S$ structure, 
with a single barrier 
at the $N/F$ interface: this is similar to the case studied in our 
analytic results. The calculation is now numerical
and fully self consistent. 

In Fig.~\ref{NFSfig1} 
we plot the total conductance $G$ as a function
of the rescaled bias voltage $E$, together with the 
spin-up and spin-down conductance contributions. 
We do so in four panels corresponding
to varying intermediate $F$ layer thickness
 $D_F$, with fixed $D_N=90$ and $D_S=180$ in our 
dimensionless units. We take the scattering
strength at the $N/F$ interface of $H_B=0.5$.
The variation in 
$D_F$ is chosen,  as in our non-self-consistent results, 
to include a thickness range that 
demonstrates a full period of the conductance's 
subgap peak behavior. The most obvious
difference between the results of the
non-self-consistent analytic calculation and 
those obtained via the numerical self-consistent procedure
 is that the latter case leads to
a varying critical bias. This has been found and 
discussed previously\cite{Moen2017} and is directly related to the 
drop in the pair potential due to the proximity effect of the pair amplitude. 

The first (upper left)
panel of Fig.~\ref{NFSfig1},  corresponds
to the situation
where the ZBC is large and the CBC is low. 
The critical bias 
itself is significantly smaller
in the self-consistent case, and there is little difference between the spin-up 
and spin-down conductance curves. 
Just as in the analytic case, this behavior is periodic with $D_F$ and occurs again near
 $D_F=30$ (not shown). 
In the second panel, 
we see the transition in the spin-split conductance, with a subgap peak in the total $G$ 
 due to the opposing 
behavior of the 
spin-up and spin-down conductance components. The spin-up conductance displays a positive concavity and a cusp feature at 
the critical bias, while the spin-down conductance displays a negative concavity with a weaker shoulder feature at the CB, similar
to those found in our analytic calculation. 
Although the critical bias conductance depends on
the spin, the CB value itself does not. This is because both
spin channels interact with the same effective pair potential, which for the single-ferromagnet system, and singlet pairing, 
 is spin independent 
since the Hamiltonian commutes with $S_z$. 

In the third panel, we see the spin-split and total conductance peak 
locations converging towards the critical bias. 
Although not shown here, the relative behavior of the spin-up and spin-down 
conductance switches abruptly for slightly different 
 values of $D_F$, with a sharp transition similar to
what is seen in the ZBC peaks of Fig.~\ref{NSCfig4} in the analytic calculation. 
In the fourth panel, we see another 
subgap conductance peak similar to that in the 
second panel but now with the spin-split conductance components switching behavior,
 as was the case in our analytic calculation. 

The spatial period we have considered corresponds to a wavelength $\pi/h$ for our 
value of $h$. 
We conclude that the self-consistent behavior of the $N/F/S$ conductance
qualitatively  displays the same 
periodic behavior as revealed by the analytic non-self-consistent 
calculations.   
However, as mentioned above, the
CB is now dependent on $D_F$ as can clearly
be seen by looking across the four panels.
In subsection~\ref{FFSspinsplit}, we will find a further, more complex behavior
by introducing an angular dependence on the system, which  affects not only the spin-split conductance peaks but also the critical bias.
As in the non-self-consistent, infinite case, the strength $H_B$ of the interfacial scattering 
at the $N/F$ interface
enhances the peak conductance behavior, 
although we do not display this feature  here. 
Furthermore, the increase in the barrier strength increases the critical bias
value, making the analytic result approximation
less inadequate in the strong-barrier case.
Thus, the existence of the subgap conductance peak 
is verified for both the analytic and numerical calculation, 
and  the peak is mostly independent of the barrier height $H_B$.

\subsection{F/N/F/S spin-split conductance}
\label{FFSspinsplit}

We now proceed, in this subsection, to the case
of major theoretical and 
practical interest, where 
we include the outer ferromagnet, realistic, finite
thicknesses and consider all interfacial barriers. We study,
in the spin valve configuration,  
the dependence on the relative orientation of the exchange fields 
of the charge transport. This angular dependence is particularly important when applied to spin valves, as any angular 
dependence in the conductance constitutes a ``valve effect'' that can be exploited. We have studied such effects in 
previous work\cite{Moen2017,wvhg} for a variety of physical parameters. In
 this subsection we 
continue to focus on the intermediate $F_2$ 
layer dependence and the oscillatory behavior of the peak conductance, which we have already noted in the $N/F/S$ 
case. 
Therefore, we concentrate  on a small subset of interfacial scattering parameters, and on the spin-split effects that arise as 
$D_{F2}$ varies. 

\begin{figure}

\vspace*{-0.4cm}
\hspace*{-1.4cm}\begin{subfigure}[b]{.65\textwidth}
\includegraphics[angle=-90,width=0.9\textwidth] {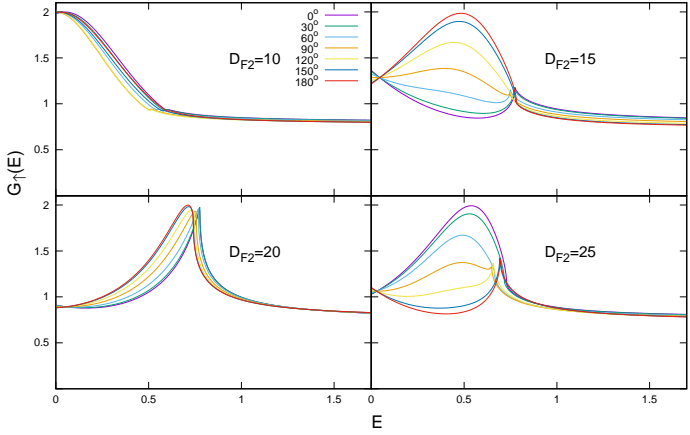} 
\vspace*{-0.6cm}
\caption{The spin-up conductance.}
\label{FNFSfig1a}
\end{subfigure}\vspace*{-0.7cm}

\hspace*{-1.4cm}\begin{subfigure}[b]{.65\textwidth}
\includegraphics[angle=-90,width=.9\textwidth] {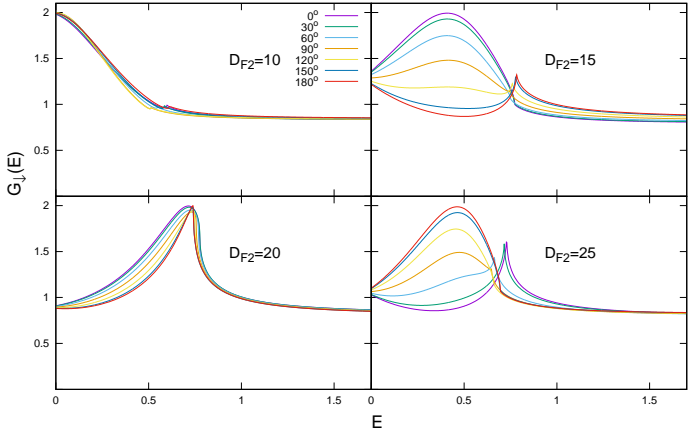} 
\vspace*{-0.6cm}
\caption{The spin-down conductance.}
\label{FNFSfig1b}
\end{subfigure}\vspace*{-0.7cm}

\hspace*{-1.4cm}\begin{subfigure}[b]{.65\textwidth}
\includegraphics[angle=-90,width=0.9\textwidth] {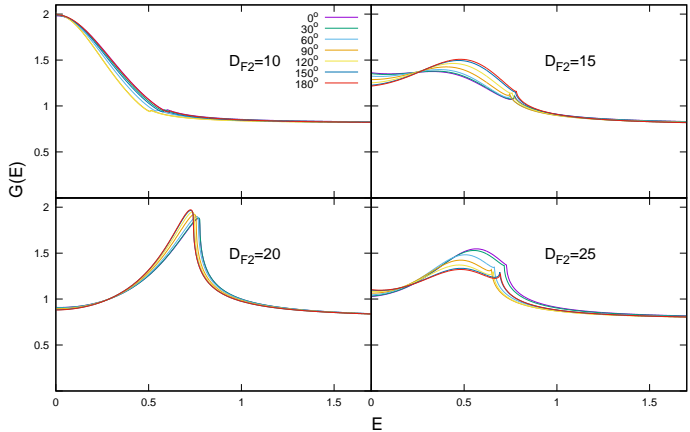} 
\vspace*{-0.6cm}
\caption{The total conductance.}
\label{FNFSfig1c}
\end{subfigure}
\caption{Spin-split and total conductance as a function of the bias and $\phi$. The above plots are for
the $F_1/N/F_2/S$ system with thicknesses $D_{F1}=30, D_N=60, D_S=180$ and a single barrier $H_B=0.5$ at the $N/F_2$ interface. The four panels in each
subfigure are for different values of the intermediate $F_2$ layer thickness in the periodic intervals of one quarter of a period.}
\label{FNFSfig1}
\end{figure}

\begin{figure}

\vspace*{-0.4cm}
\hspace*{-1.4cm}\begin{subfigure}[b]{.65\textwidth}
\includegraphics[angle=-90,width=0.9\textwidth] {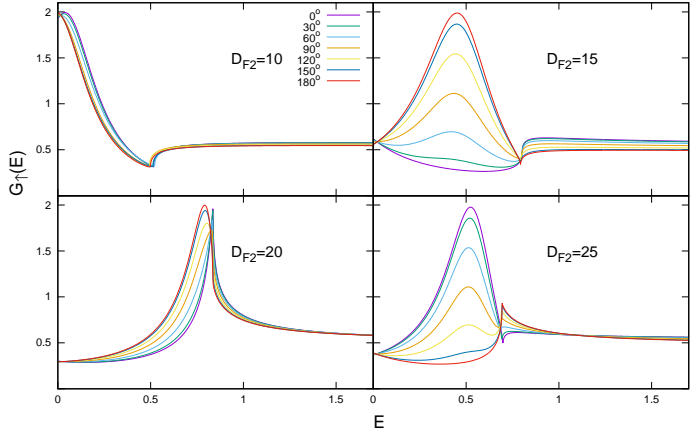} 
\vspace*{-0.6cm}
\caption{The spin-up conductance.}
\label{FNFSfig2a}
\end{subfigure}\vspace*{-0.7cm}

\hspace*{-1.4cm}\begin{subfigure}[b]{.65\textwidth}
\includegraphics[angle=-90,width=.9\textwidth] {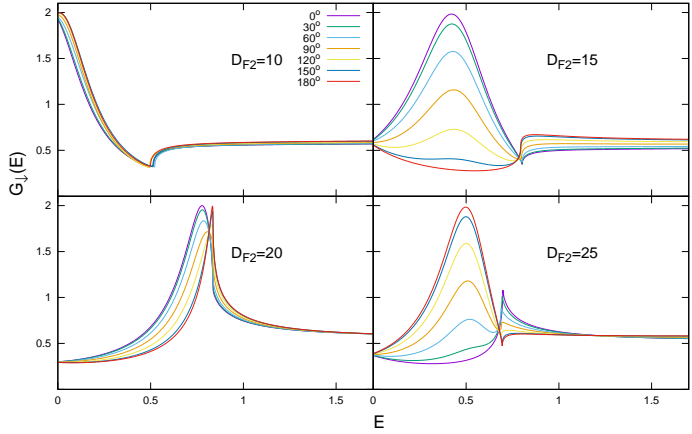} 
\vspace*{-0.6cm}
\caption{The spin-down conductance.}
\label{FNFSfig2b}
\end{subfigure}\vspace*{-0.7cm}

\hspace*{-1.4cm}\begin{subfigure}[b]{.65\textwidth}
\includegraphics[angle=-90,width=0.9\textwidth] {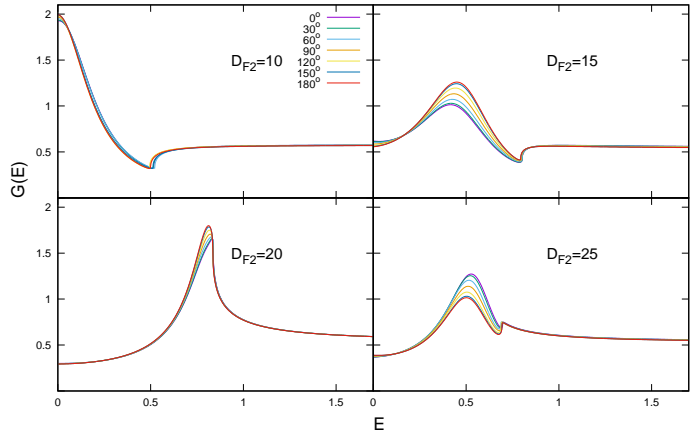} 
\vspace*{-0.6cm}
\caption{The total conductance.}
\label{FNFSfig2c}
\end{subfigure}
\caption{Spin-split and total conductance as a function of the bias and $\phi$. The above plots are for
the $F_1/N/F_2/S$ system with thicknesses $D_{F1}=30, D_N=60, D_S=180$ and a single barrier $H_B=0.9$ at the $N/F_2$ interface, significantly higher than that used in Fig.~\ref{FNFSfig1}.
The four panels in each
subfigure are for different values of the intermediate $F_2$ layer thickness as
in Fig.~\ref{FNFSfig1}.}
\label{FNFSfig2}
\end{figure}

\begin{figure}

\vspace*{-0.4cm}
\hspace*{-1.4cm}\begin{subfigure}[b]{.65\textwidth}
\includegraphics[angle=-90,width=0.9\textwidth] {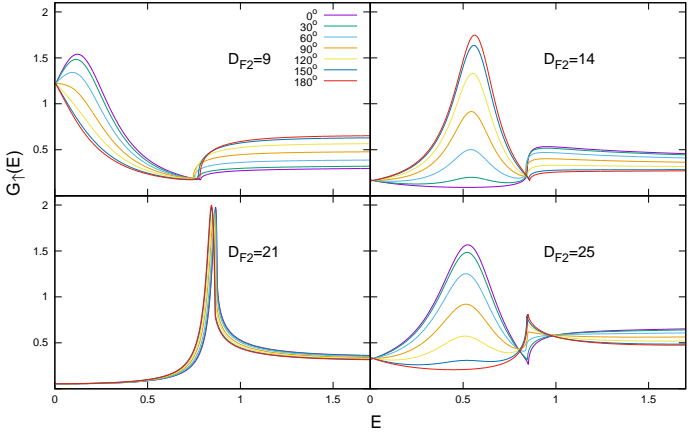} 
\vspace*{-0.6cm}
\caption{The spin-up conductance.}
\label{FNFSfig3a}
\end{subfigure}\vspace*{-0.7cm}

\hspace*{-1.4cm}\begin{subfigure}[b]{.65\textwidth}
\includegraphics[angle=-90,width=.9\textwidth] {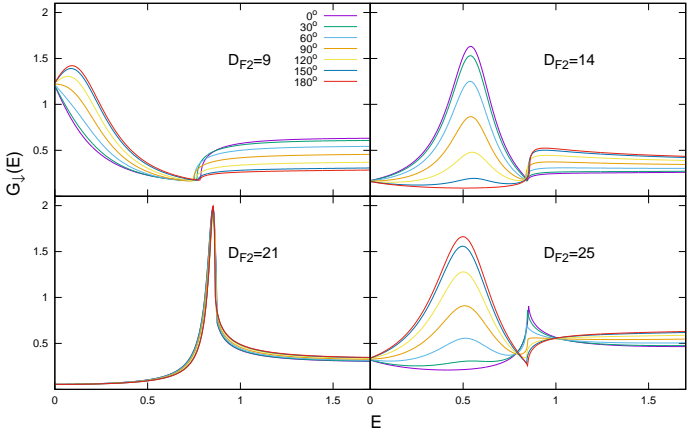} 
\vspace*{-0.6cm}
\caption{The spin-down conductance.}
\label{FNFSfig3b}
\end{subfigure}\vspace*{-0.7cm}

\hspace*{-1.4cm}\begin{subfigure}[b]{.65\textwidth}
\includegraphics[angle=-90,width=0.9\textwidth] {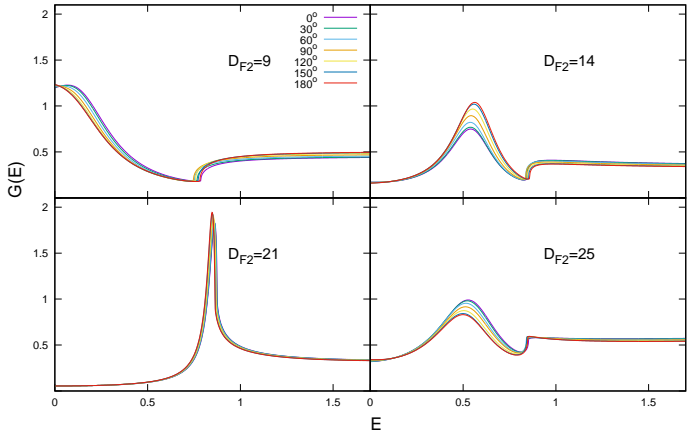} 
\vspace*{-0.6cm}
\caption{The total conductance.}
\label{FNFSfig3c}
\end{subfigure}
\caption{Spin split and total conductance as a function of the bias and $\phi$. The above plots are for
the $F_1/N/F_2/S$ system with thicknesses $D_{F1}=30, D_N=60, D_S=180$ and realistic barriers $H_{Bi}$ equal to $0.5$, $0.5$, and $0.3$ at the $F_1/N$, $N/F_2$, and
$F_2/S$ interfaces respectively.
The four panels in each subfigure are for different values of the intermediate $F_2$ layer thickness in the  intervals of one quarter period.}
\label{FNFSfig3}
\end{figure}

In Figs.~\ref{FNFSfig1} and ~\ref{FNFSfig2} 
we plot the spin-split and the total 
conductance as a function of 
the misalignment angle $\phi$ for a single
interfacial barrier, located  at the $N/F_2$ interface, as 
was done in Fig.~\ref{NFSfig1}. Introducing this 
barrier best exhibits the behaviors of the peak conductance 
that can occur. 
We will later include one full set of  barriers,  in 
Fig.~\ref{FNFSfig3}. 
We will also in this case
be plotting the  dependence of the  conductances
on the misalignment angle $\phi$. Therefore, we subdivide 
each figure into three parts: (a) The spin-up conductance, (b) the spin-down conductance, and (c) the total conductance. In each
of these parts the panels correspond
to different values of $D_{F2}$, as indicated.

In Fig.~\ref{FNFSfig1} 
we plot the 
mentioned quantities 
as a function of the bias and $\phi$ for a moderate barrier value $H_{B2}=0.5$. 
Here, we see that the 
spin-up and spin-down components (Figs.~\ref{FNFSfig1a} 
and \ref{FNFSfig1b} respectively)   are 
highly dependent on the relative angle of magnetization. 
It is obviously no coincidence 
that the spin-up conductance very closely resembles that
of the spin-down conductance for supplementary angles.
Much of this resemblance is due to 
the change in $\phi$ being accounted for, in large
part, by a purely mathematical rotation 
of the spin-split conductance 
as given by Eq.~(\ref{Gspineq}). 
Thus, it is seen that 
under a rotation by an angle $\theta$,
$G_\uparrow(\theta)=G_\downarrow(\pi-\theta)$ and vice versa.
The angular dependence of each spin 
component closely resembles a combination of $\phi=0$ of the spin-up and spin-down conductance, rotated into
the respective $\phi$ basis via Eq.~(\ref{Gspineq}) for $\theta\rightarrow\phi$.
For the same reason, it should  be no surprise that a subgap peak 
in the spin-split conductance is found near $\phi=90^\circ$, since
this can be largely described by a combination of the spin-up and spin-down conductances, as is the case with the total conductance.
However, not all the differences in the features between the spin-up and spin-down conductances can be explained by this rotation,
and a true angular dependence exists that is different for each component of the spin-split conductance. This yields a
much more complex angular dependence in the total conductance (Fig.~\ref{FNFSfig1c}). 

In Fig.~\ref{FNFSfig1a}, we plot the spin-up conductance. 
We see a considerable spread in the critical bias. 
The angular dependence is relatively weak in the first panel and
becomes much stronger in the other three. In the second panel, the CB increases 
for angles greater than $90^\circ$ and 
decreases for angles less than $90^\circ$. 
In the third and fourth panels, we see the opposite: the CB decreases for 
$\phi>90^\circ$ and increases for $\phi<90^\circ$. Recall
 that in the $N/F/S$ case we saw the spin-up and spin-down conductance 
swap behavior in panels two and four of Fig.~\ref{NFSfig1}, with a transition occurring 
in panels one and three. Similarly, we see here the CB behavior also 
making this transition in its angular dependence. The cusp and shoulder behavior of the CBC is not qualitatively
changed by the introduction 
of the second ferromagnet. 
We also see 
a split in the ZBC. 

To complement the previous subplot,
we next display, in Fig.~\ref{FNFSfig1b}, the spin-down conductance. As mentioned above, 
the behavior of this quantity is similar to that of the spin-up 
conductance, but with an  angular dependence shifted by $\pi$. 
The angular dependence of the ZBC and the CBC 
have dramatically changed, with opposite behavior. The ZBC no longer has a crossing point in the low bias regime. In effect, the 
introduction of the second ferromagnet takes the crossing node of the ZBC seen in the 
$N/F/S$ case (see Fg.~\ref{NFSfig1}) and moves it to the right 
in the spin-up conductance and to the left in the spin-down conductance. 
The CBC for the spin-down conductance experiences broadening, 
in direct opposition 
of the spin-up conductance, as can be seen best in 
the right hand panels (panels two and four). 
The angular dependence of the CB also broadens in 
these two panels. In panel three, we see the CB 
values move closer together and reverse the order of their angular dependence. 
This is explained by  
the spin-up conductance and spin-down conductance being
 at different phases in their $D_F$ periodicity. 
In panel three, we see the critical bias (and the overall conductance) behavior transition in its angular dependence from that
of panel two to that of panel four.
From the CB features plotted, we see, due to the spin-valve effect, that the spin-down conductance is slightly advanced in its phase, 
while the spin-up conductance lags behind. 

Finally, in the last panel set, Fig.~\ref{FNFSfig1c}, we  analyze the 
overall impact of the second ferromagnetic layer by plotting
 the total conductance, which can
then be compared  to that in Fig.~\ref{NFSfig1}. The
total $G$, as
given by  Eq.~(\ref{spinsplitcond}), is not, 
unlike in the $N/F/S$ case, simply the 
average of the spin-up and spin-down conductances, because the
outer electrode 
$F_1$ is populated with a majority of spin-up electrons:  the 
total conductance is now weighted more heavily towards the spin-up 
value. Therefore we see an angular dependence in the total 
conductance that is more reminiscent of that  of the spin-up conductance. 
This can be seen in the similar CB angular dependence as well as
the ZBC and CBC dependences. 
The combination of spin-up and spin-down conductance leaves us with a smaller
subgap peak in the total $G$, for all angles. 
Generally, we see a 
significantly reduced angular dependence when compared to the spin-split conductance, except for the ZBC and the CBC. 
We also see that the cusp and shoulder CB features are less pronounced. 
There is a crossing point in panels two and four as 
we saw with the spin-up conductance, however this is not an exact ``node" as the conductance does not cross over at precisely the same bias 
for all angles. In panel three, we actually see a monotonically increasing peak 
conductance, even though  
neither the spin-up nor spin-down conductance feature this monotonic behavior. 
In this transition, the phase difference of the 
spin-up and spin-down oscillations with respect to $D_{F2}$ has a greater impact on the CBC behavior of the total $G$ 
than for thicknesses such as in panels two and four.

In Fig.~\ref{FNFSfig2} 
we study the impact of the barrier strength on the spin-split
conductance  by increasing the 
parameter barrier value used in 
Fig.~\ref{FNFSfig1} 
from $H_{B2}=0.5$ to $H_{B2}=0.9$, the value used
in Fig.~\ref{NSCfig2}. 
We have found in previous work\cite{Moen2017} that an increase in barrier strength 
can lead to a decrease in angular dependence, particularly for the critical bias. We have also found above, in the $N/F/S$ 
case, that increasing the barrier strength can enhance the subgap conductance peak behavior. Below we analyze the combined 
effect that this change makes on our results.

In Figs.~\ref{FNFSfig2a} and \ref{FNFSfig2b} we plot the spin-up and spin-down 
conductances, respectively, for this 
 larger barrier value.  
We see the supplementary
angle  relation 
 in the $\phi$
behavior of the spin-split conductance. 
We also see an 
angular dependence arise in the CB and the ZBC, as we did 
in Figs.~\ref{FNFSfig1a} and \ref{FNFSfig1b}, but this angular dependence 
is much smaller: this reflects the overall suppression of
the proximity effects by the higher barrier. Furthermore, 
the difference between the spin-up and spin-down conductances (besides the switching of 
conductance behavior to supplementary angles) is greatly diminished. 
We do see a small broadening in the angular dependence of 
the CB, as well as a better defined crossing point. 
This leads to a total 
conductance that has, in the zero bias
and critical bias regions,  little angular dependence, as we see in Fig.~\ref{FNFSfig2c}. 
However, the subgap conductance peak still maintains a strong
 angular dependence, rivaling that of the $H_{B2}=0.5$ case. 
This is because much of the angular dependence here comes from the difference in spin-up and spin-down electron populations 
emanating from the $F_1$ layer, in which the large difference 
between spin-up and spin-down conductance counteracts the decrease 
in angular dependence of the other conductance features. We also note that for higher
 barriers this subgap peak is more pronounced. 
The reason is twofold: the increased difference between the spin-up and spin-down conductances creates a large peak in the total 
$G$, as we saw in Fig.~\ref{NSCfig2}, and the decrease in the angular dependence of the CB provides less overlap, which 
prevents the hybridizing of the cusp and shoulder CB behaviors, and makes the drop-off sharper near the CB.

We now examine in Fig.~\ref{FNFSfig3} 
the realistic case where there are barriers at all three interfaces:
 $H_{B1}=H_{B2}=0.5$, $H_{B3}=0.3$ for 
the $F_1/N$, $N/F_2$, and $F_2/S$ interfaces respectively. These values
are  likely a good approximation to real experimental 
conditions as there are unavoidable interfacial defects even in the best
prepared  samples of heterostructures. The introduction 
of the $F_2/S$ barrier can slightly flatten the subgap peak conductance feature
because a barrier at that interface reduces the proximity effect.
However, this effect remains small, 
and when coupled with the moderate $F/N$ barriers, it
leaves the conductance with well defined peaks, as we will see below. 
Having two interfacial barriers, 
particularly with similar values, can produce  resonance 
effects\cite{Moen2017} 
in the low bias conductance for certain thicknesses. 
There are a large number of parameter choices that can affect
 the conductance 
features in a variety of ways, but what is of interest here
 is how robust the oscillatory subgap peak behavior is.

In Fig.~\ref{FNFSfig3a} we plot the spin-up conductance for these 
barrier values, and in Fig.~\ref{FNFSfig3b} the 
spin-down conductance. The first thing to observe is the slight 
change in the $D_{F2}$ values displayed in the first three
panels from thhe values used 
in  previous figures. We show, in the first panel,
thickness
values  closest to the transition where the subgap 
conductance peak is now
closest to zero bias. The ZBC peak conductance in panel one is very 
sensitive to small changes in $D_{F2}$, and it is important to try to
tune precisely to that value. 
This sensitivity indicates that these barriers can have a large 
impact on the phase of the oscillatory behavior of 
the peak conductance. We see in our plot
a large angular 
dependence at the peak bias, which is  slightly greater than zero,
but  very little angular dependence on the ZBC. Also, 
 the angular dependence in 
the subgap bias range is large, as we found
in the single barrier,
$H_{B2}=0.9$ case (see Fig.~\ref{FNFSfig2}). 
The high bias conductance ($E>1$) now displays a large 
angular dependence in panels one, two and four, but 
this dependence is much smaller in panel 
three  where the subgap peak transitions at the CB. 
Examining the spin-up and 
spin-down conductances, we see greater broadening with $\phi$ in the 
CB and the CBC than we did in Fig.~\ref{FNFSfig2} in panels two and four. 
In panels one and two, we see some slight phase advancing/lagging, but the other conductance features are quite similar. These 
transitional panels have peaks which are sharper than in the previous 
results, and the ZBC peak is lower in value.

In Fig.~\ref{FNFSfig3c} we plot the total conductance. Its behavior
 is similar to that of Fig.~\ref{FNFSfig2c} but there are
some key differences. Despite the $F_2/S$ barrier, the subgap conductance peak is sharper in panels two and four. 
Furthermore there is a larger, more noticeable angular dependence in the CB, as well as in the high-bias conductance. 
The ZBC, however, has a smaller angular dependence, but does still 
feature a small-bias crossing point before the subgap
peak conductance. In panel three we see again an angular dependence in the CBC similar to that in Fig.~\ref{FNFSfig1c}. Overall, 
the salient point is that the subgap peak behavior is not
only still present, but in fact more pronounced, with a large angular 
dependence in the peak conductance in panels two and four. This peak conductance is oscillatory with $D_{F2}$, with only a slight change in phase 
resulting from the introduction of 
 realistic barriers. This robust angular dependence of the peak conductance can potentially be exploited, 
as the subgap conductance peak leads to an angularly dependent  change
 in the excess currents at high biases. 

\section{Conclusions} 
\label{conclusions} 

We have analyzed here the spin-split
conductance in $F_1/N/F_2/S$ spin valve systems using  numerical, self-consistent
methods. We have also considered $N/F/S$ systems using also an approximate
but analytic method.
We have found a peak in the subgap conductance that is periodic with the
intermediate $F$ layer thickness. This peak conductance is due 
to the separate behavior 
of the 
contributions to the total conductance from
incoming spin-up and spin-down electrons. We collectively 
call these contributions the spin-split conductance. 
Our results
show that the subgap conductance peak position oscillates
between the zero bias and the critical bias values as $D_F$ varies. 
We find that at least one spin
band conductance has a maximum close to $G=2$ in our natural
units at a single bias value in the subgap region, near where the opposite spin
band has a minimum. At this  subgap bias, we find a
pronounced peak in the total conductance due to the spin-up and spin-down conductances 
being very different at this bias, while they 
converge in the ZBC and CBC. In
$N/S$ systems with moderate or tunneling barriers, a peak in the conductance occurs 
at the critical bias before decreasing to normal conductance\cite{btk}. In the $F/S$
case we now see a second, subgap peak that is robust to  interfacial 
scattering, with large angular dependence in the $F_1/N/F_2/S$ spin valve. The subgap peak
and the ZBC resonance peak we find are qualitatively similar to 
the ``double peak spectra'' and the ``zero peak spectra'' seen in the tunneling conductance
measurements, reported in Ref.~[\onlinecite{Bernardo2015}], made
via scanning tunneling spectroscopy in $S/F$ structures with a non-homogeneous (Holmium) ferromagnet. 
Our theoretical work, however, focuses on spin-valve structures. 

Our spin valve results are numerical. It is usually difficult to gain
physical intuition from purely numerical results. In an effort to gain
additional intuitive understanding, we have
used an approximate, non-self-consistent,
analytic approach for an infinite $N/F/S$ structure. 
We examined the origin of the spatial periodic behavior
by examining the ZBC and CBC as a function of the thickness of the $F$ layer. 
In both cases, we found that the 
periodic spatial dependence is due to the interaction of the spin
dependent plane wave amplitudes in the ferromagnet, 
which leads to a  wavelength of $\pi/h$ in the conductance peak ($h$ is the exchange field
of the intermediate ferromagnet). 
The location of
the subgap conductance peak was found to oscillate 
 between zero bias and the CB. 
The spin-split conductance as a function of bias consequently
 switches behavior between spin components with changes in $F$ layer thickness, 
transitioning across the CBC and ZBC peak conductances.  
In the analytic non-self-consistent approximation, there is no change to the critical bias itself, which is incorrect.
We also established the effect of the interfacial barrier heights,
on the conductance features. The subgap conductance peak is only weakly dependent on the barrier strength. Furthermore,
when this peak occurs in the middle of the subgap region, the 
ZBC and CBC values decrease at a faster rate with increasing barrier strength, 
leaving a more pronounced subgap peak at higher barrier strengths. 

Turning to the numerical self-consistent results
 for the $F_1/N/F_2/S$ spin valve,  we find
the same periodic effects, with an additional dependence of the spin-split conductance 
on the CB.  This is also 
found for a finite $N/F/S$ structure. This dependence on the CB is reduced by high barriers.
We analyze the dependence of the spin-split
conductance on the angle $\phi$ between the internal
exchange fields in the magnets 
$F_1$ and $F_2$, in the $F_1/N/F_2/S$ system. 
Part of the angular dependence of the spin-split conductance in this system 
can be attributed to rotations in spin space 
(see Eq.~(\ref{Gspineq})) but since (except at $\phi=0$ and $\phi=\pi$), $S_z$
does not commute with the Hamiltonian, we find that this dependence  on $\phi$  is beyond
 that arising from a choice of spin quantization axis.
This affects the CB, the CBC, and the ZBC in different ways for the spin-up and spin-down components, causing broadening
of the CBC peaks and CB values, as well as shifting the crossing 
points where $G$ is approximately equal for all angles $\phi$. 
From our analysis we conclude that the phase of the 
periodic $D_{F2}$ dependence is 
angularly dependent for
both components of the spin-split conductance: the thickness
at which the $G$ behavior
transitions depends on $\phi$. 
There is a general shift in the spin-split conductance's bias 
dependence, in opposite directions for each component, with a nodal point,
 located at zero bias in the $N/F/S$ system,
 shifting to higher bias values for spin-up and to lower ones for spin-down.
The end result is that the total conductance has a complex angular 
dependence, where the subgap peak becomes less prominent, as the relative shift 
of the spin-split conductance means that each component's respective (at supplementary 
angles) extrema are no longer aligned, leaving their combination (i.e. the total 
conductance) more smeared, and the other
conductance features less pronounced. 
Nevertheless, a subgap conductance peak with a very strong angular dependence
remains in the $F_1/N/F_2/S$ structure.
This angular dependence is protected by the subgap peak, which
 does not diminish strongly with increasing barriers.

The subgap conductance peak, due to the spin-split conductance, 
is an important
and prominent feature that can be exploited in future superconducting spintronic devices.
One of our primary goals here
has been to determine and improve the efficacy of a superconducting spin 
valve in which the valve effect is defined by the angular dependence of 
the exchange fields.
The subgap peak is well defined when the interface between the 
superconductor and the valve is reasonably clean, even when the interfacial 
scattering within the valve is non-negligible.
Although this can lead to very low angular dependence when the peak conductance 
is at zero bias or at the CB, the angular dependence 
is large and robust against interfacial
scattering for definite values of the intermediate ferromagnetic layer 
thickness. By tuning the thickness 
to one of these intermediate values, 
a valve effect in the excess current can be attained, as
we see then a very large angular dependence in the spin-split and total conductance. 
This would have a considerable effect on the quality
of such spin valve devices.

\acknowledgments  The authors thank I.N. Krivorotov (University
of California, Irvine) for many
illuminating discussions on the
experimental issues. They are very grateful to Chien-Te Wu
(National Chiao Tung University) for many helpful
discussions on aspects of this
problem. This work was supported in part by DOE grant No. DE-SC0014467


\begin{thebibliography}{00}

\bibitem{tsyzu} E. Tsymbal and I. \v{Z}uti\'{c}, {\it Handbook on spin transport and magnetism}, CRC Press, Boca Raton, Florida (2012).
\bibitem{Bhatti2017} Sabpreet Bhatti, Rachid Sbiaa, Atsufumi Hirohata, Hideo Ohno, Shunsuke Fukami, S.N. Piramanayagam, {\it Spintronics based random access memory: a review}, Materials Today, ISSN 1369-7021 (2017).
\bibitem{Buzdin2005} A. I. Buzdin, Rev. Mod. Phys. {\bf 77}, 935 (2005).
\bibitem{igor} I. \v{Z}uti\'{c}, J. Fabian, and S. Das Sarma, \rmp {\bf 76}, 323 (2004).
\bibitem{esch} M. Eschrig, Rep. Prog. Phys. {\bf 78}, 104501 (2015).
\bibitem{zkhv} J. Zhu, I.N. Krivorotov, K. Halterman and O.T. Valls,
\prl {\bf 105}, 207002 (2010).
\bibitem{wvhg} C-T Wu, O.T. Valls and K. Halterman, \prb {\bf 90}, 054523,
(2014).
\bibitem{kami} T. Yu. Karminskaya, A.A. Golubov, and M. Yu. Kupryanov, \prb
{\bf 84}, 064531 (2011).
\bibitem{transistor} Nevirkovets, I. P., Chernyashevskyy, O., Prokopenko, G. V., Mukhanov, O. A. and Ketterson, J. B. Superconducting-ferromagnetic transistor. IEEE Trans. Appl. Supercond. {\bf 24}, 1800506 (2014).
\bibitem{birge} E. C. Gingrich, B. M. Niedzielski, J. A. GLick, Y. Wang, D. L. Miller, R. Loloee, W. P. Pratt Jr, N. O. Birge, Nature Physics {\bf 12}, 564-567 (2016). 
\bibitem{Moen2017} E. Moen, O.T. Valls, \prb {\bf 95}, 054503 (2017).
\bibitem{Moen2018} E. Moen, O.T. Valls, \prb {\bf 97}, 174506 (2018).

\bibitem{Buzdin1990} Buzdin, A. I., and M. Y. Kuprianov, Pisima Zh. Eksp. Teor. Phys. {\bf 52}, 1089-1091 [JETP Lett. {\bf 52}, 487-491 (1990)].
\bibitem{Halterman2002} K. Halterman and O. T. Valls, \prb  {\bf 66}, 224516 (2002).
\bibitem{demler} E. A. Demler, G. B. Arnold, and M. R. Beasley, \prb {\bf 55}, 15174 (1997).
\bibitem{ryaz2001} V. V. Ryazanov, V. A. Oboznov, A. Yu. Rusanov, A. V. Veretennikov, A. A. Golubov, and J. Aarts
Phys. Rev. Lett. 86, 2427 (2001).
\bibitem{birge2018} Bethany M. Niedzielski, T. J. Bertus, Joseph A. Glick, R. Loloee, W. P. Pratt, Jr., and Norman O. Birge
Phys. Rev. B {\bf 97}, 024517 (2018).
\bibitem{Bernardo2015} A. Di Bernardo, S. Diesch, Y. Gu, J. Linder, G. Divitini, C. Ducati, E. Scheer, M.G. Blamire, and J.W.A. Robinson, Nat. Commun. {\bf 6}, 8053 (2015).

\bibitem{hoprl}C.T. Wu, O.T. Valls, and K. Halterman, \prl {\bf 108}, 117005 (2012).
\bibitem{berg86} F.S. Bergeret,  A.F Volkov, and K.B. Efetov, \prl {\bf 86}, 3140 (2001);
\prb {\bf 68}, 064513 (2003). 
\bibitem{hbv} K. Halterman, P. Barsic and O.T. Valls, \prl {\bf 99} 127002 (2007).
\bibitem{bvh} P.H. Barsic, O.T. Valls and K. Halterman, \prb {\bf 75}, 104502
(2007).
\bibitem{zdravkov2013} V. I. Zdravkov, J. Kehrle, G. Obermeier, D. Lenk, H.-A. Krug von Nidda, C. Miller, M. Yu. Kupriyanov, A. S. Sidorenko, S. Horn, R. Tidecks, and L. R. Tagirov \prb {\bf 87}, 144507 (2013).
\bibitem{berezinskii} V. L. Berezinskii, JETP Lett.  {\bf 20}, 287 (1975).
\bibitem{bvermp} F.S. Bergeret,  A.F Volkov, and K.B. Efetov, \rmp {\bf 77}, 1321 (2005).
\bibitem{kalcheim} Y. Kalcheim, O. Millo, A. Di Bernardo, A. Pal and J.W. Robinson,
\prb {\bf  92}, 060501 (2015).
\bibitem{singh} A. Singh, S. Voltan, K. Lahabi, and J. Aarts, 
Phys. Rev. X {\bf 5}, 021019 (2015).
\bibitem{ha2016}K.Halterman and M. Alidoust, Phys. Rev. B {\bf 94} 064503 (2016).
\bibitem{alejandro} A. A. Jara, C. Safranski, I. N. Krivorotov, C.-T. Wu. 
A. N. Malmi-Kakkada,
O. T. Valls, and K. Halterman, \prb {\bf 89}, 184502 (2014).

\bibitem{johnson1998} M. Johnson, \prb {\bf 58}, 9635 (1998). 
\bibitem{johnson2001} M. Johnson, Physica E {\bf 10}, 472 (2001). 
\bibitem{Andreev} A. F. Andreev, Sov. Phys. JETP {\bf 19}, 1228 (1964).
\bibitem{linder2009} J. Linder, T. Yokoyama, and A. Sudb\o, \prb {\bf 79}, 224504 (2009).
\bibitem{visani}C. Visani, Z. Sefrioui, J. Tornos, C. Leon, J. Briatico, 
M. Bibes, A. Barth\'{e}l\'{e}my, J. Santamar\'{i}a, and Javier E. Villegas,
Nature Phys. {\bf 8}, 539 (2012).
\bibitem{niu} Z. P. Niu, Europhys. Lett. {\bf 100}, 17012 (2012).
\bibitem{ji} Y.-Q. Ji, Z.-P. Niu, C.-D. Feng, and D.-Y. Xing, Chinese Phys. Lett. {\bf 25}, 691 (2008)

\bibitem{degennes} P. G. de Gennes, {\it Superconductivity of Metals and Alloys} (Addison-Wesley, Reading, MA, 1989).
\bibitem{btk} G. E. Blonder, M. Tinkham, and T. M. Klapwijk, \prb {\bf 25}, 4515 (1982).


\bibitem{bagwell}  P.F. Bagwell, \prb {\bf 49}, 6841 (1993). 
\bibitem{sols2} F. Sols and J. Ferrer, \prb {\bf 49}, 15913 (1994). 
\bibitem{sols} J. Sanchez-Canizares and F. Sols, \prb {\bf 55}, 531 (1997).
\bibitem{baym} G. Baym and L.P. Kadanoff, Phys. Rev. {\bf 124},  287 (1961).

\end{thebibliography}
\end{document}